\documentclass[12pt]{article}
\usepackage{amsfonts,amssymb}
\usepackage{mathrsfs}

\newcommand{\DD}{{\mathscr{D}}}
\newcommand{\R}{{\mathbb{R}}}
\newcommand{\Z}{{\mathbb{Z}}}

\newcommand{\I}{{\mathbb{I}}}

\newcommand{\beq}{\begin{equation}}
\newcommand{\eeq}{\end{equation}}
\newcommand{\bea}{\begin{eqnarray}}
\newcommand{\eea}{\end{eqnarray}}
\newcommand{\ra}{\rightarrow}

\newcommand{\cd}{\partial}
\newcommand{\less}{\backslash}
\newcommand{\wt}{\widetilde}

\newcommand{\su}{{\mathfrak{su}}}

\newcommand{\tr}{{\rm tr}\, }

\newcommand{\vol}{{\rm vol}}

\newcommand{\etavec}{\mbox{\boldmath{$\eta$}}}
\newcommand{\tauvec}{\mbox{\boldmath{$\tau$}}}

\begin{document}

\title{A pure Skyrme instanton}
\author{
J.M. Speight\thanks{E-mail: {\tt speight@maths.leeds.ac.uk}}\\
School of Mathematics, University of Leeds\\
Leeds LS2 9JT, England
}

\date{}
\maketitle

\begin{abstract}
The nuclear Skyrme model is considered in the extreme limit where
the nucleon radius tends to infinity. In this limit only the
Skyrme term in the action is significant. The model is then
conformally invariant in dimension 4, and supports an
instanton solution which can be constructed explicitly.
The construction uses the
conformal invariance and a certain
symmetry reduction to reduce the model to
the static $\phi^4$ model
in one dimension. The $\phi^4$ kink solution gives the radial profile
of the instanton, the kink position zero-mode corresponding to the
instanton width. 
\vspace*{0.3cm}\newline
\end{abstract}

\maketitle

The Skyrme model is a low-energy effective theory of nuclear physics
possessing a Lie group valued field $U:\R^{3+1}\ra SU(2)$.  The action is 
usually 
taken to be
\beq\label{skyact}
S=\int_{\R^{3+1}}\frac{F_\pi^2}{16}\tr(L^\mu L_\mu^\dagger)+
\frac1{32e^2}\tr([L_\mu,L_\nu][L^\mu,L^\nu])
\eeq
where $L_\mu=U^{-1}\cd_\mu U$ is the left invariant current, Greek
indices run over $0,1,2,3$, and we have given spacetime the
signature $\displaystyle{\begin{array}{cccc}+&-&-&-\end{array}}$. 
This model possesses topological solitons, 
labelled topologically
by their class $B\in\pi_3(SU(2))\cong \Z$. These are thought to model
light atomic nuclei, $B$ being identified with 
the number of nucleons in the nucleus. 
This physical interpretation is used to set the values of the
coupling constants $F_\pi$ and $e$. $F_\pi$ is the pion decay constant,
and the nucleon radius is proportional to $(eF_\pi)^{-1}$ \cite{adknapwit}.

The purpose of this letter is to show that in the extreme limit
$F_\pi e\ra 0$, where only the second term in $S$ is
important, the model possesses {\em instantons}, that is finite
action solutions of the Euclideanized model on $\R^4$. We call this
limit, which corresponds physically to the limit
of large nucleon radius, the pure Skyrme model. 
The instantons can be constructed explicitly by working within 
a rotationally equivariant ansatz for which the model reduces to the
static $\phi^4$ model (in one dimension). The radial profile of the 
instanton is directly related to the $\phi^4$ kink profile. The instanton
is labelled topologically by its class in $\pi_4(SU(2))\cong\Z_2$.
Hence it coincides with the anti-instanton (the concatenation of two
instantons is null-homotopic).

The action of
interest is $S=\frac{1}{2e^2}E_4$ where
\beq
E_4=\frac1{16}\int_{\R^{4}}\tr([L_\mu,L_\nu][L_\mu,L_\nu]).
\eeq
The notation $E_4$ signifies that we are thinking of the functional
as the potential energy of a static field on $\R^4$, the subscript
4 denoting that the
energy density is quartic in spatial derivatives.
For our purposes it is convenient to use Manton's geometric
formulation of the Skyrme term \cite{man}. Recall this makes sense
for maps $U:M\ra N$ between any pair of Riemannian manifolds
$(M,g)$, $(N,h)$, 
interpreted as physical space and target space respectively. In
our case $M=\R^4$ with the Euclidean metric,
and $N=SU(2)\cong S^3$ with the round metric of unit radius. 
Associated to any $U:M\ra N$ there is a symmetric 
$(1,1)$ tensor $\DD$
on $M$, that is, a self-adjoint linear map $\DD_p:T_pM\ra
T_pM$ for each $p\in M$, called the strain tensor, defined by
\beq
g_p(X,\DD_p Y)=h_{U(p)}(dU_p X, dU_pY)
\eeq
for all $X,Y\in T_pM$.
Manton showed that the Skyrme term $E_4$ is, in this language,
\beq
E_4=\frac12\int_M (\tr\DD)^2-\tr(\DD^2).
\eeq
Under a conformal change in the metric on $M$, $g\mapsto \wt{g}=e^{2f}g$
where $f\in C^\infty(M)$, the strain tensor of a given configuration
$U$ transforms as 
$\DD\mapsto \wt{\DD}=e^{-2f}\DD$, and the volume form on $M$ transforms
as $\vol\mapsto \wt{\vol}=e^{\dim_\R M f}\vol$. So the energy
functional $E_4$ is conformally
invariant if (and only if) $M$ has dimension $4$, 
the case of interest here. In this sense, the pure Skyrme model is
similar to Yang-Mills theory.

We will seek critical points of $E_4$ within a radially symmetric ansatz,
defined as follows. Split $\R^4\less\{0\}$ into a family of concentric
three-spheres labelled by radial coordinate $r$. Identify each
$S^3$ with $SU(2)$ in the usual way
(see equation (\ref{s3su2}), below). Let $g_{SU(2)}$ be the usual 
bi-invariant metric on $SU(2)$ (of unit radius). Then the Euclidean
metric on $\R^4\less\{0\}$ is
\beq
g=dr^2+r^2g_{SU(2)}=e^{2s}(ds^2+g_{SU(2)})
\eeq
where $s:=\log r$. Since 
$\R^4\less\{0\}$ is conformal to $\R\times SU(2)$ and $E_4$ is conformally
invariant, we can equally well solve the model on $\R\times SU(2)$, with
the product metric (we must check that the decay of $U$ as $|s|\ra \infty$
is sufficiently fast for the energy to converge as $r\ra 0$ and 
$r\ra \infty$). 
Now consider fields of the form
\beq\label{ansatz}
U:\R\times SU(2)\ra SU(2),\quad
U(s,q)=q\eta(s)q^{-1}
\eeq
where $\eta:\R\ra SU(2)$ is any curve in $SU(2)$ satisfying
the boundary conditions
 $\eta(-\infty)=-\I$, $\eta(\infty)=\I$. As we shall see in the next 
paragraph, for fixed $s$, $U(s,q)$ lies on a two-sphere in
$SU(2)$ for all $q\in SU(2)$. 
Fields within this ansatz are precisely those which are invariant under the 
left action of $SU(2)$ defined by
\beq\label{su2act}
U(s,q)\stackrel{V}{\mapsto} VU(s,V^{-1}q)V^{-1}.
\eeq
This action maps $\DD(s,q)\stackrel{V}{\mapsto}\DD(s,V^{-1}q)$ and hence 
leaves $E_4$ invariant. Hence, by
the principle of symmetric criticality, critical points of
the restriction of $E_4$ to fields of this form are solutions of
the full variational problem for $E_4$ \cite{pal}, so the
ansatz (\ref{ansatz}) is guaranteed to be consistent with the
variational equation for $E_4$. 

We claim that, provided the curve $\eta$ avoids the poles $\eta=\pm\I$
and satisfies the boundary conditions $\lim_{s\ra\pm\infty}=\pm\I$,
the corresponding field $U(s,q)=q\eta(s)q^{-1}$
(or, more precisely, its unique continuous extension to $S^4$)
lies in the nontrivial class of $\pi_4(SU(2))\cong \Z_2$. To see this,
consider for each fixed $s\in\R$ the map $U(s,\cdot):SU(2)\ra SU(2)$.
We can identify the target $SU(2)$ with the unit three-sphere in $\R^4$ using
\beq\label{s3su2}
(a_0,a_1,a_2,a_3)\leftrightarrow
a_0\I+ia_1\tau_1+ia_2\tau_2+ia_3\tau_3=
\left[
\begin{array}{cc}a_0+ia_3&-a_1+ia_2\\a_1+ia_2&a_0-ia_3\end{array}
\right],
\eeq
where $\tau_1,\tau_2,\tau_3$ are the Pauli spin matrices.
Then $U(s,\cdot)$ is a Hopf map from
$S^3$ onto the two-sphere obtained by intersecting
$S^3$ with the hyperplane $a_0=\eta_0(s)$, 
where 
\beq
\eta(s)=:\eta_0(s)\I+i\etavec(s)\cdot\tauvec.
\eeq
 Hence the extension of
$U(s,q)$ to $S^4$
is topologically a suspension of the Hopf map $S^3\ra S^2$. Now the
Hopf map generates $\pi_3(S^2)$, and
suspension induces a surjective homomorphism $\pi_3(S^2)\ra\pi_4(S^3)$
by the Freudenthal suspension theorem \cite{hat},
so the extension of $U$ is not null-homotopic. This conclusion agrees with
the results of Williams \cite{wil} who considered the topology of similar
maps $S^4\ra S^3$.

It remains to substitute the ansatz (\ref{ansatz}) into $E_4$ and
vary the curve $\eta$.
Let us introduce an orthonormal frame for $M=\R\times SU(2)$ consisting
of the vector field $e_0=\frac{\cd\: }{\cd s}$ and the left invariant
vector fields $e_1,e_2,e_3$ on $SU(2)$ whose values at $\I$ coincide with
$i\tau_1,i\tau_2,i\tau_3$. Note that $e_1,e_2,e_3$ are {\em twice}
the usual left invariant vector fields $\theta_1,\theta_2,\theta_3$
on $SU(2)$, so our orthonormal frame on $M$ is
\beq\label{basis}
e_0=\frac{\cd\: }{\cd s},\quad
e_1=2\theta_1,\quad
e_2=2\theta_2,\quad
e_3=2\theta_3.
\eeq
Relative to this
basis the strain tensor of
a field of form (\ref{ansatz}) has matrix representative
\beq
\DD=\left[\begin{array}{cc}
\dot{\eta}_0^2+|\dot{\etavec}|^2&-2(\etavec\times\dot{\etavec})^T\\
-2(\etavec\times\dot{\etavec})& 4(|\etavec|^2\I_3-\etavec\otimes\etavec^T)
\end{array}\right],
\eeq
where $\dot{}$ denotes differentiation with respect to $s$ and $\times$
is the $\R^3$ vector product.
Hence the energy functional to be minimized is
\beq
E_4=2\pi^2\int_{-\infty}^\infty ds\, \left\{
(1+|\etavec|^2)\dot{\eta}_0^2+|\etavec|^2|\dot{\etavec}|^2+4|\etavec|^4
\right\}
\eeq
subject to the constraint $\eta_0^2+|\etavec|^2\equiv 1$. 
Clearly this functional is invariant under the discrete symmetry
\beq\label{discrete}
(\eta_0,\eta_1,\eta_2,\eta_3)\mapsto (\eta_0,-\eta_1,-\eta_2,\eta_3).
\eeq
Hence, by the principle of symmetric criticality again, we may seek
solutions which are fixed by this symmetry, that is, we can restrict
attention to curves of the form $(\eta_0(s),0,0,\eta_3(s))$ with
$\eta_0^2+\eta_3^2\equiv 1$. For such curves $E_4$ reduces to 
\beq
E_4=8\pi^2\int_{-\infty}^\infty ds\, \left\{\frac12\dot\eta_0^2+(1-\eta_0^2)^2
\right\}.
\eeq
Note that this is precisely the potential energy of the $\phi^4$ model
in one dimension.
We seek solutions of this model interpolating between $\eta_0=-1$ and
$\eta_0=1$. There is a one-parameter family of such solutions, the kinks, 
parametrized by position on the line $s_0\in\R$,
\beq\label{kink}
\eta_0(s)=\tanh\sqrt2(s-s_0).
\eeq
Since the $\phi^4$ kink is known to have finite energy, it is immediate
that the decay of $U(s,q)=q\eta(s)q^{-1}$ as 
$|s|\ra\infty$ is fast enough to ensure 
finite total $E_4$. In fact $E_4=32\sqrt2\pi^2/3$.
Transforming back to polar coordinates on $\R^4\less\{0\}$, we find that
\beq\label{profile}
U(r,q)=q\left\{\frac{1}{(r/r_0)^{2\sqrt2}+1}\left[\begin{array}{cc}
(r/r_0)^{2\sqrt2}+2i(r/r_0)^{\sqrt2}-1&0\\
0&(r/r_0)^{2\sqrt2}-2i(r/r_0)^{\sqrt2}-1
\end{array}\right]\right\}q^{-1},
\eeq
where $r_0=e^{s_0}$. Note that
the kink position parameter $s_0$ becomes the width of the instanton 
$r_0$ on
$\R^4$, a free parameter of the solution, reflecting
the conformal invariance of $E_4$. Note also that the
instanton is smooth away from the origin but only $C^1$ at the origin. 
Alternatively, we can interpret $U(s,q)$ as a globally smooth, spatially
homogeneous, instanton on (Euclideanized) spacetime $\R\times S^3$.

By the
usual Bogomol'nyi argument applied to the $\phi^4$ model, we know that
the instanton  (\ref{profile})
minimizes $E_4$ among all {\em equivariant}
maps in its homotopy class. It is an open question whether the instanton
is a true minimum of $E_4$ among all maps in its 
homotopy class: perhaps there is a globally
smooth minimizer outside the equivariant ansatz.

As a check on our construction, we can verify directly that the mapping
constructed satisfies the field equation for action $E_4$. This is
\beq\label{fec}
\cd_\alpha([L_\beta,[L^\beta,L^\alpha]])=0
\eeq
where $x^\alpha$ are Cartesian coordinates on $\R^4$ and 
$L_\alpha=U^{-1}\cd_\alpha U$
is (as before) the left-invariant current \cite{mansut}. While one
could compute the left hand side of (\ref{fec}) by brute force using, for
example, Maple, it is more satisfactory to complete the calculation
by hand, and for this it is useful
to re-write equation (\ref{fec}) in coordinate independent language.

Let $\mu$ be the left-invariant Maurer-Cartan form on $SU(2)$, that is,
the $\su(2)$ valued 1-form which assigns to any vector $X\in T_qSU(2)$
the element $\hat{X}\in\su(2)=T_{\I}SU(2)$ whose left translate by $q$
coincides with $X$. Then $L_\alpha=(U^*\mu)(\cd_\alpha)$ 
(so we are interpreting the left-invariant current as a 1-form on $M$, 
the pullback by
the field $U$ of the Maurer-Cartan form). Further,
we define a $\su(2)$ valued 1-form $\nu$ on $M$ by
\beq\label{nudef}
\nu(X)=\sum_{i}[U^*\mu(e_i),[U^*\mu(e_i),U^*\mu(X)]]
\eeq
where $\{e_i\}$ is any orthonormal frame on $M$, and $X$ is any vector on 
$M$. Then (\ref{fec}) may be rewritten
\beq\label{feci}
\delta\nu=0
\eeq
where $\delta$ is the coderivative on $\su(2)\otimes T^*M$. (More concretely,
$\delta\nu=\sum_{a=1}^3(-*d*\nu_a)i\tau_a$ where 
$\nu=\sum_a\nu_a(i\tau_a)$, so $\nu_a$ are real 1-forms on
$M$.) This gives an alternative geometric
formulation of the variational problem: the 1-form $\nu$, constructed
from $U^*\mu$, must be coclosed.

Now, for fields within the equivariant ansatz
(\ref{ansatz}), we have
\beq\label{mc}
U^*\mu=q\eta^\dagger\{\dot{\eta}\, ds+\sum_{a=1}^3\frac{i}{2}[\tau_a,q]
\sigma_a\}q^\dagger
\eeq
where $\{\sigma_a\}$ are the 1-forms dual to $\{\theta_a\}$. 
Specializing further, for the instanton
$\eta(s)=\eta_0(s)\I_2+i\eta_3(s)$, $\eta_0^2+\eta_3^2\equiv 1$ and 
$\eta_0(s)$ satisfies the Bogomol'nyi equation 
\beq\label{bog}
\dot{\eta}_0=\sqrt{2}(1-\eta_0^2), 
\eeq
so that (\ref{mc}) simplifies to
\beq\label{mc2}
U^*\mu=-i\eta_3q\{\sqrt{2}\tau_3\, ds+(\eta_3\tau_1-\eta_0\tau_2)\sigma_1+
(\eta_0\tau_1+\eta_3\tau_2)\sigma_2\}q^\dagger.
\eeq
From this it quickly follows that
\beq
\nu=i\eta_3^3q\{32\sqrt{2}\tau_3\, ds+24(\eta_3\tau_1-\eta_0\tau_2)\sigma_1
+24(\eta_0\tau_1+\eta_3\tau_2)\sigma_2\}q^\dagger.
\eeq
It remains to check that $\nu$ is coclosed. For this we note that
\cite{ura}
\beq
\delta\nu=-\sum_i(\nabla_{e_i}\nu)(e_i)=-\sum_i\{e_i[\nu(e_i)]
-\nu(\nabla_{e_i}e_i)\}
\eeq
where $\nabla$ is the Levi-Civita connexion on $M$. Now each vector field
in the frame (\ref{basis}) has geodesic integral curves, so 
$\nabla_{e_i}e_i=0$ for all $i$ for this frame, so
\bea
\delta\nu&=&-\sum_ie_i[\nu(e_i)]\nonumber\\
&=&
-\frac{\cd\:}{\cd s}\left(32\sqrt{2}i\eta_3^3q\tau_3q^\dagger\right)
-2\theta_1\left[48i\eta_3^3q(\eta_3\tau_1-\eta_0\tau_2)q^\dagger\right]
-2\theta_2\left[48i\eta_3^3q(\eta_0\tau_1+\eta_3\tau_2)q^\dagger\right]
\nonumber\\
&=&32\sqrt{2}i\, 3\eta_3^2\dot{\eta}_3q\tau_3q^\dagger
-96i\eta_3^3q[\frac{i}{2}\tau_1,\eta_3\tau_1-\eta_0\tau_2]q^\dagger
-96i\eta_3^3q[\frac{i}{2}\tau_2,\eta_0\tau_1+\eta_0\tau_3]q^\dagger
\nonumber\\
&=&192i\eta_0\eta_3^3q\tau_3q^\dagger
-96i\eta_3^3\eta_0q\tau_3q^\dagger-96i\eta_3^3\eta_0q\tau_3q^\dagger
=0
\eea
where we have once again used the Bogomol'nyi equation (\ref{bog}). 
This completes the check that the claimed
instanton satsifies the field equation.

It may seem that the limit $F_\pi e\ra 0$ is rather artificial. 
Certainly, if a multiple of the quadratic term
\beq
E_2=\int_{\R^4}\frac12\tr(L_\mu L_\mu^\dagger)=\int_{\R^4}\tr\DD
\eeq
is added to $E_4$ then no instanton solution on $\R^4$
is possible, by Derrick's theorem \cite{der}. The point is that, 
under a scaling variation
$U(x)\mapsto U(\lambda x)$, where $\lambda>0$ is a positive constant,
$
E_2\mapsto\lambda^{-2} E_2$,
while
$E_4\mapsto E_4$, so a nonconstant configuration can always lose energy
by shrinking.
The modern viewpoint of the Skyrme model, however, is to think of
it as an expansion in derivatives of a more fundamental action
coming from QCD, truncated at the
quartic term, so that sextic and higher terms in the action density
are neglected \cite{wit}. 
The main reason for truncating at the quartic term is 
that
this gives the simplest theory which evades
Derrick's theorem in dimension 3 and hence allows
topological solitons. 
In principle, there is no reason why terms sextic (and higher) in
spatial derivatives should not also be included. One possibility which has
been examined in detail in $\R^{3+1}$ \cite{flopie}  is the term
\beq
E_6=\int_{\R^4}\tr\left\{
[L_\mu,L_\nu] [L_\nu, L_\lambda] [L_\lambda, L_\mu]\right\}.
\eeq
Generalized Skyrme models of this type evade Derrick's theorem in
dimensions 3 and 4, so they could quite possibly support both Skyrmions 
and instantons. Note that $E_2$ and $E_6$ are, like $E_4$, invariant
under the $SU(2)$ action (\ref{su2act}), so the ansatz (\ref{ansatz})
works in this more general setting too. Conformal invariance is lost,
however, as is the reduction to $\phi^4$ theory,
and the differential equation for the curve $\eta(r)$
does not appear to be analytically tractable. Nevertheless, one would hope
that
instanton solutions exist within this ansatz, and could be constructed
at least numerically.

\subsection*{Acknowledgements} The author thanks Steffen Krusch for useful
correspondence.

\end{document}